\documentclass[aps,amssymb,amsmath,12pt,floatfix]{revtex4}
\usepackage{subfigure}
\usepackage{graphicx}
\usepackage{rotating}
\usepackage{multirow}
\usepackage{color}

\begin{document}

\title{Order statistics inference for describing topological coupling and mechanical symmetry
breaking in multidomain proteins}

\author{Olga Kononova$^{1,2}$, Lee Jones$^3$ and V. Barsegov$^{1,2}$}
\thanks{Corresponding author; phone: 978-934-3661; fax: 978-934-3013;
Valeri\_Barsegov@uml.edu}
\affiliation{$^1$Department of Chemistry, University of Massachusetts, Lowell, MA 01854, United States \\
$^2$Moscow Institute of Physics and Technology, Moscow Region, Russia 141700\\
$^3$Department of Mathematical Sciences, University of Massachusetts, Lowell, MA 01854, United States}

\date{\today}


\begin{abstract}

\noindent
Cooperativity is a hallmark of proteins, many of which show a modular architecture comprising 
discrete structural domains. Detecting and describing dynamic couplings between structural 
regions is difficult in view of the many-body nature of protein-protein interactions. By utilizing
the GPU-based computational acceleration, we carried out simulations of the protein forced 
unfolding for the dimer $WW$$-$$WW$ of the all-$\beta$-sheet $WW$ domains used as a model 
multidomain protein. We found that while the physically non-interacting identical protein 
domains ($WW$) show nearly symmetric mechanical properties at low tension, reflected, 
e.g., in the similarity of their distributions of unfolding times, these properties become 
distinctly different when tension is increased. Moreover, the uncorrelated unfolding transitions 
at a low pulling force become increasingly more correlated (dependent) at higher forces. Hence, 
the applied force not only breaks ``the mechanical symmetry'' but also couples the physically
non-interacting protein domains forming a multi-domain protein. We call this effect ``the topological 
coupling''. We developed a new theory, inspired by Order statistics, to characterize protein-protein 
interactions in multi-domain proteins. The method utilizes the squared-Gaussian model, but it can also 
be used in conjunction with other parametric models for the distribution of unfolding times. The formalism 
can be taken to the single-molecule experimental lab to probe mechanical cooperativity and domain 
communication in multi-domain proteins.

\end{abstract}
\maketitle


\section{INTRODUCTION}

Mechanical forces play an important role in life processes, and many biological molecules are subjected 
to forces. There are numerous examples from biology, where mechanical forces play an essential role in 
a physiological context. Tensile forces acting on cells or bacteria originate from the dragging forces 
imposed by the fluid flow \cite{AFMGaub}. A function of a biological motor might result in the generation 
of large pulling force. For example, active force from myosin generates substantial mechanical stress on 
structures of the sarcomere \cite{SarcomereGautel,ActinKasza}. Mechanical forces can also be generated 
within the biological system. For example, leukocytes that patrol the blood flow in search of pathogens, 
need to generate internal force to squeeze into in and pass through connective tissues \cite{SelectinMcEver}. 
In addition to biochemical stimuli, cellular processes involving the cytoskeleton can be modulated in 
response to external force \cite{ActinKasza}. 
 
Structural arrangements of proteins have evolved in response to the selection pressure from biological 
forces. The three-dimensional structures often exhibit a modular architecture composed of 
discrete structural regions. These include semi-static structures such as actin filaments and microtubules, 
and flexible structures such as muscle protein titin and fibrin clot \cite{FilaminsNakamura,TitinTskhovrebova,Weisel1}. 
Titin, a giant $1 \mu m$ protein formed by linked immunoglobulin and fibronectin domains defines 
the structure and elasticity of muscle sarcomere \cite{TitinTskhovrebova,SarcomereGautel}. Proteins 
that form or interact with the extracellular matrix (ECM) have modular or multidomain architecture 
\cite{CadherinLeckband,ECMRamage}. These include fibronectin fibrils, which form meshworks 
around the cell, and integrins, which link the cell external and intracellular environment \cite{MTDWang}. 
Fibrin polymerization in blood results in formation of branched network called a fibrin clot, which 
must sustain large shear stress due to blood flow \cite{Weisel1,Weisel2}. Protein 
shells of plant and animal viruses (capsids) are often made of multiple copies of a single structural unit 
(capsomer). For example, the capsid of Cowpea Chlorotic Mottle Virus is an icosahedral shell comprised of 
$180$ copies of a single ($190$ amino acid) protein \cite{CCMVSpeir}.

Physical properties of proteins have adapted to biological forces. In ECM assembly, fibronectin subunits 
change conformations from compact to extended. The extent of assembly is controlled by cell contractility 
through actin filaments \cite{FibronectinSingh}. Hence, tissue tension regulate matrix assembly. Virus 
capsids should be stable enough to protect encapsulated material (DNA or RNA), yet, unstable to release 
the material when invading their host cells \cite{VirologyRoos}. Hence, mechanical properties of virus 
capsids are important factors in viruses' survival in the extracellular environment and cell infectivity. 
Cell adhesion and migration rely on reversible changes in mechanical properties of cells. Spatial distribution 
of internal tension in cell remodeling, derived from dynamic regulation of contractile actin-myosin networks, 
correlates cell shape and movement \cite{ActinLieleg,ActinKasza}. In addition, many proteins have evolved 
to act as ``force sensors'' to convert tension-induced conformational changes into biological signal 
\cite{MTDWang}. For example, tissue deformation during mechanical stimulation alters the conformation of 
ECM molecules. Filamins involved in mechanotransduction alter their structure in response to external 
tension \cite{FilaminSutherlandSmith}. Proper communication between structural regions in multidomain 
proteins is important for regulation. For example, dynamic coupling between the regulatory domain and the 
ligand-binding domain in P-selectins is implicated in formation of force-activated bonds \cite{CatchHertig} 
(``catch bonds'' \cite{Selectin}) with their ligands (PSGL-1). 

Single-molecule experimental techniques such as Atomic Force Microscopy (AFM) \cite{AFMGaub,AFMRief,Brujic1} 
and optical trap \cite{OptTweezBlock,AFMRief}, have made it possible to study life processes at the level 
of individual molecules. These experiments, which utilize mechanical force to unfold proteins or to dissociate 
protein-protein complexes, have made it possible to probe the unfolding or unbinding transitions one at a 
time \cite{ReceptLigandLitvinov}. In the context of protein forced unfolding, grabbing the molecule at 
specific positions allows one to select specific region(s) of the molecule and to define the unfolding 
reaction coordinate. Mechanical manipulation continues to provide a unique approach to quantify the unfolding 
transitions in terms of the measurable quantities - the unfolding forces (constant velocity or force-ramp 
experiment) and the unfolding times (constant force or force-clamp experiment). Consider a 
force-clamp experiment on a multimeric protein $D_n$ formed by head-to-tail connected identical domains 
($D$'s). The protein is subjected to the external pulling force, and each unfolding transition causes 
an increase in the chain length, $\Delta Y$. As a result, the total length of the 
polypeptide chain $Y$ shows the characteristic pattern of stepwise increases 
($\Delta Y$'s), which mark sequential unfolding transitions in protein domains. 

The goal of statistical analysis and modeling of force spectroscopy data is to obtain accurate 
information about the physical characteristics of protein domains. Yet, due to inherent limitations in the 
current experimental resolution, there is no direct way of attributing the times of transitions to the 
protein domains or structural regions where these transitions have occurred. We only have partial or 
incomplete observation of the system. Said differently, it is not possible to determine which specific domain 
has unfolded at any time since any domain can unfold at any given time. Consider the simplest case of the 
dimer $WW$$-$$WW$ of the $WW$ domain, presented in Fig.~1. There are two possible scenarios of 
unfolding. In the first pathway, the first domain ($WW_1$) unfolds first and the second domain ($WW_2$) 
unfolds second. In the second pathway, the order of unfolding is reversed. In an experimental measurement, 
what is being recorded is the first unfolding time ($t_{1:2}$) and the second unfolding time ($t_{2:2}$) 
in a sequence of two observations, and it is impossible to tell which domain has unfolded first or second. 
Hence, the first unfolding times ($t_{1:2}$) and the second unfolding times ($t_{2:2}$) involve 
contributions from the first domain $WW_1$, which unfolds at time $t_1$, and the second domain $WW_2$, which 
unfolds at time $t_2$. Can we obtain domain-specific information from the experimental data?

The first step is to understand the nature of the random variables measured. In a constant force experiment 
on an $n$-domain protein ($D_1$$-$$D_2$$-$$\ldots$$D_n$), individual domains unfold one 
after another but any domain can unfold at any given time. The observed unfolding times are ordered, i.e. they 
comprise a set of ordered time variates, also known as Order statistics \cite{Gumbel,David}. Hence, what is being 
measured are the ``time-ordered data'' $t_{1:n},t_{2:n},\ldots,t_{n:n}$, where $t_{r:n}$ is the $r$-th unfolding 
time ($r$$=$$1,2,\ldots,n$) (in a sequence of $n$ observations). These are different from the (hidden) 
``parent data'' $t_1,t_2,\ldots,t_n$ with $t_i$ being the unfolding time of the $i$-th domain 
($D_i$, $i$$=$$1,2,\ldots,n$), which contain information about the individual protein domains 
($D_1,D_2,\ldots,D_n$). Hence, the question becomes - can we solve an inverse problem, namely, can 
we perform the inference of the parent distributions of unfolding times from the distributions of (observed) 
ordered unfolding times? 

The $r$-th order statistic is characterized by the cumulative distribution functions (cdf's) $P_{r:n}(t)$,
 and the probability density functions (pdf's) $p_{r:n}(t)$ of the $r$-th unfolding time, 
$r$$=$$1,2,\dots,n$, in a sequence of $n$ observations (for $n$ domains). The $r$-th order statistic cdf 
$P_{r:n}(t)$ is the probability that the $r$-th unfolding time $t_r$ does not exceed 
$t$, i.e., $P_{r:n}(t)$$=$$Prob(t_r\leq t)$, and the $r$-th order statistic pdf is 
$p_{r:n}(t)$$=$$dP_{r:n}(t)/dt$ \cite{Gumbel}. In our example for the dimer $WW$$-$$WW$,
 the first unfolding time ($t_{1:2}$) and the second unfolding time ($t_{2:2}$) both carry information 
about the unfolding times for the first domain $WW_1$ ($t_1$) and for the second domain $WW_2$ ($t_2$).
 In general, because the cdf $P_{r:n}(t)$ and the pdf $p_{r:n}(t)$
 of the $r$-th order statistic depend on the parent cdf $P_i(t)$ and pdf $p_i(t)$ 
($i$$=$$1,2,\ldots,n$), it is possible, at least in principle, to resolve $P_i(t)$ and $p_i(t)$ from the cdf 
and pdf of the order statistics, $P_{r:n}(t)$ and $p_{r:n}(t)$.

In our recent papers \cite{OS1,OS2}, we used Order statistics to solve the inverse problem for 
the independent identically distributed ({\em iid}) random variables and for the independent non-identically 
distributed ({\em inid}) random variables. The formalism presented in these papers 
can be used to describe the physically non-interacting identical protein domains ($D$'s) forming a multimeric 
protein $D$$-$$D$$-$$\ldots$$-$$D$ ({\em iid} case), and the non-interacting non-identical domains ($D_i$'s, 
$i$$=$$1,2,\ldots,n$) forming a multi-domain protein $D_1$$-$$D_2$$-$$\ldots$$-$$D_n$ ({\em inid} case) \cite{OS1}. 
We have designed rigorous statistical tools for assessing the independence of the parent forced unfolding 
times ($t_i$) and the equality of the parent pdfs of unfolding times ($p_i(t)$) from the observed ordered 
unfolding times ($t_{r:n}$) \cite{OS2}. These statistical tests can be utilized to classify the parent unfolding 
times and to detect correlated unfolding transitions in multi-domain proteins using experimental force spectroscopy
data. We have also extended Order statistics approach to describe the unfolding forces 
measured in the constant-velocity (force-ramp) experiments \cite{OS3}. Order statistics inference 
methods have been applied successfully to inverse problems involving, e.g., partialy observed queuing systems 
\cite{Lee1,Lee2,Larson}. 

Here we take a step further and present a new theory inspired by Order statistics, to solve 
the inverse problem for the dependent identically distributed ({\em did}) random variables and for the dependent 
non-identically distributed ({\em dnid}) random variables, using a squared-Gaussian parametric model of the 
distributions of unfolding times. The developed formalism can be used to describe the mechanical behavior of 
interacting (coupled) identical protein domains in a multimeric protein $D$$-$$D$$-$$\ldots$$-$$D$ ({\em did} 
case), and interacting non-identical domains in a multidomain protein $D_1$$-$$D_2$$-$$\ldots$$-$$D_n$ ({\em dnid} 
case). In the next Section, we describe our method. We focus on the order statistics pdf's, since these statistical measures can be easily estimated by constructing the histograms of unfolding times. We establish 
a relashionship between the order statistics pdf's and the parent pdf's. In Section III, we describe the Self 
Organized Polymer (SOP) model of the all-$\beta$-sheet $WW$ domain and Langevin simulations of the forced 
unfolding of the dimer $WW$$-$$WW$ used as a model system (Fig.~1). The {\em in silico} experiments mimic 
single-molecule measurements {\em in vitro} \cite{Fg1,Fg2,ProtofilamentsDima,C2DomainDima}. In section IV, 
we perform a direct statistical analysis of the simulation output for dimer $WW$$-$$WW$ and describe the effects 
of ``mechanical symmetry breaking'' and ``topological coupling''. In Section V, we compare several statistical 
measures of the ''parent data'' for each $WW$ domain - the distribution of unfolding times, 
the average unfolding time, the standard deviation, and the skewness of distribution, and Pearson correlation 
coefficient, with the same measures obtained by applying Order statistics inference to the ``time-ordered data''. 
We discuss our results in Section VI.


\section{ORDER STATISTICS INFERENCE}

Consider a vector of order statistics ${\bf T^\prime}$$=$$[t_{1:n}, t_{2:n}, \dots, t_{n:n}]^\dagger$, 
in which the entries obtained in a single measurement correspond to the $1$-st unfolding time 
($t_{1:n}$), $2$-nd unfolding time ($t_{2:n}$), $\dots$, and $n$-th unfolding time ($t_{n:n}$), the symbol 
``dagger'' ($\dagger$) represents vector transpose ($t_{r:n}$ denotes the $r$-th unfolding time out of $n$ 
times for a protein $D_1$$-$$D_2$$-$$\ldots$$-$$D_n$ of $n$ domains $D_1,D_2,\ldots,D_n$). This same observation 
${\bf T^\prime}$ can be also obtained by rearranging the components of another vector 
${\bf T}$$=$$[t_1,t_2,\dots,t_n]^\dagger$ in the order of increasing time variates, which represents the 
unfolding times of the $1$-st domain  $D_1$ ($t_1$), $2$-nd domain $D_2$ ($t_2$), $\dots$, and 
$n$-th domain $D_n$ ($t_n$) of the same multidomain protein $D_1$$-$$D_2$$-$$\ldots$$-$$D_n$. This is 
because any domain can unfold at any given time with a non-zero probability and, hence, there are multiple 
unfolding scenarios. Hence, on the one hand, we observe the ``time-ordered data'' ($t_{r:n}$), and on the 
other hand, we have (hidden) "parent data" ($t_i$). We call this transformation map ${\bf G}$ and write 
${\bf T^\prime}$$=$${\bf G}({\bf T})$$=$$[t_{1:n}, t_{2:n}, \dots, t_{n:n}]^\dagger$, where 
$t_{r:n}$ is the $r$-th smallest component of vector ${\bf T}$.

Let vector ${\bf T^\prime}$ have the joint pdf $p_{\bf T^\prime}(t_{1:n}, t_{2:n}, \dots, t_{n:n})$ 
and let vector {\bf T} have the joint pdf $p_{\bf T}(t_1, t_2, \dots t_n)$. We seek to establish a 
relationship between the pdf of the ordered data, $p_{\bf T^\prime}(t_{1:n}, t_{2:n}, \dots, t_{n:n})$, 
and the pdf of the parent data, $p_{\bf T}(t_1, t_2, \dots t_n)$. Let us first assume, that 
vector ${\bf T}$ can be written as ${\bf T}$$=$$[t_1,t_2,\dots,t_n]^\dagger$$=$$[x_1^2,x_2^2,\dots,x_n^2]^\dagger$, 
where ${\bf X}$$=$$[x_1, x_2, \dots, x_n]^\dagger$ is an arbitrary random vector sampled from the Gaussian 
distribution
\begin{equation} \label{eq:xPDF}
p_{\bf X}(x_1, x_2, \dots, x_n) = \frac{1}{(2\pi)^{n/2}\sqrt{|{\bf {\bf \Sigma}}}|}
e^{-\frac{1}{2}({\bf X}-{\bf \mu})^{\dagger}{\bf \Sigma}^{-1}({\bf X}-{\bf \mu})}
\end{equation}
where ${\bf \mu}$$=$$[\mu_1, \mu_2, \dots, \mu_n]^\dagger$ is the vector of the mean, and 
${\bf \Sigma}$$=$$[\sigma_{ij}]_{i,j=1}^n$ is the covariance matrix of vector ${\bf X}$. Here, $|{\bf \Sigma}|$ 
denotes the determinant of matrix ${\bf \Sigma}$. We use squared-Gaussian distribution since it yields one-sided 
exponential tail behavior and a non-zero skewness (typical of Gamma-process), but unlike Gamma distribution, 
squared-Gaussian is simple to describe correlations of the data.

We next obtain the joint probability density function of the parent data ${\bf T}$ in terms of 
the random variable ${\bf X}$, by using the following general transformation formula for the joint pdf of the 
multivariate random variable:
\begin{equation}\label{eq:gtf}
p_{{\bf T}}(t_1, t_2, \dots, t_n)=\sum_{l}{p_{{\bf X}}(x^l(t_1, t_2, \dots, t_n))}||{\bf J}^l(t_1, t_2, \dots, t_n)||
\end{equation}
which is valid for a smooth many-to-one mapping from real space $R^n$ to real space $R^n$ with the non-zero Jacobian 
of the inverse mapping. In Eq. (\ref{eq:gtf}), the sum runs over all inverse branches, $x^l(t_1, t_2, \dots, t_n)$, 
with $l$ being the $l$-th branch, and $||{\bf J}^l||=\Bigl|\det\left[\frac{\partial x_i^l({\bf t})}{\partial t_j}\right]_{i,j=1}^n\Bigr|$ 
is the absolute value of the Jacobian of the $l$-th inverse branch. The total number of inverse branches for the 
transformation $t_i$$=$$x_i^2$ ($i$$=$$1,2,\dots,n$) is $2^n$, and they can be written as $x_i^l$$=$$s_i^l\sqrt{t_i}$, 
$s_i^l$$=$$\pm1$, $l$$=$$1,2,\dots,2^n$ ($i$$=$$1,2,\dots,n$). The Jacobian matrix for this transformation is 
given by a diagonal matrix, with the $i$-th diagonal entry given by $J_i^l$$=$$\frac{s_i^l}{2\sqrt{t_i}}$. 
The determinant of this matrix is $|{\bf J}^l|$$=$$\prod_{i=1}^n{\frac{s_i^l}{2\sqrt{t_i}}}$. 
Substituting the expressions for $p_{\bf X}$ and $|{\bf J}^l|$ in Eq. (\ref{eq:gtf}), we arrive
at the joint pdf of the parent data expressed in terms of the pdf of vector ${\bf X}$:
\begin{equation}\label{eq:parentPDF}
p_{{\bf T}}(t_1, t_2, \dots, t_n) = \frac{1}{(2\pi)^{n/2}\sqrt{|{\bf \Sigma}|}} 
\cdot \frac{1}{2^n\sqrt{t_1\cdot t_2 \cdots t_n}} \sum_{\bf s}{e^{-\frac{1}{2}
({\bf s}{\bf \sqrt{T}}-{\bf \mu})^{\dagger}{\bf \Sigma}^{-1}({\bf s}{\bf \sqrt{T}}-{\bf \mu})}}
\end{equation}
where the sum varies over all vectors of signs ${\bf s}$$=$$[s_1, s_2, \dots, s_n]^\dagger$ with the 
entries $s_i$$=$$\pm$$1$, and ${\bf s}{\bf \sqrt{T}}$$=$$[s_1\sqrt{t_1},s_2\sqrt{t_2},\dots,s_n\sqrt{t_n}]^\dagger$ 
is the vector of inverse transformation written in terms of the components ($i$$=$$1,2,\dots,n$).

The final step is to obtain the pdf of the time-ordered data, $p_{\bf T^\prime}(t_{1:n},t_{2:n},\dots,t_{n:n})$, 
in terms of the pdf of the parent data using Eq. (\ref{eq:gtf}). Consider the map ${\bf G}$, and use this map in 
Eq. (\ref{eq:gtf}). Since ${\bf G}$ orders the components of vector ${\bf T}$ to produce the order
statistics vector ${\bf T^\prime}$$=$$[t_{1:n},t_{2:n},\dots,t_{n:n}]^\dagger$, the inverse branches 
of ${\bf G}$ are all possible permutations, $\kappa$, of the components of vector ${\bf T^\prime}$. 
In the case of $n$ domains, there are a total of $n!$ possible rearrangements of $n$ components and, 
hence, $n!$ of the inverse branches of ${\bf G}$. Since the Jacobian of the permutation transformation 
is either $+1$ or $-1$, the formula connecting the joint pdf of the order statistics data and the joint 
pdf of the parent data reads: 
\begin{eqnarray}\label{eq:orderStatPDF}
p_{{\bf T}^\prime}(t_{1:n}, t_{2:n}, \dots, t_{n:n}) 
& = & \sum_{\kappa}{p_{{\bf T}}(t_{\kappa(1):n}, t_{\kappa(2):n}, \dots, t_{\kappa(n):n})}\nonumber\\
& = & \frac{1}{(2\pi)^{n/2}\sqrt{|{\bf \Sigma}|}} \cdot \frac{1}{2^n\sqrt{t_{1:n}\cdot t_{2:n}\cdots t_{n:n}}} 
\sum_{\kappa}{\sum_{\bf s}{e^{-\frac{1}{2}({\bf s}{\bf \sqrt{T_\kappa}}-{\bf \mu})^{\dagger}
{\bf \Sigma}^{-1}({\bf s}{\bf \sqrt{T_\kappa}}-{\bf \mu})}}}
\end{eqnarray}
where the symbolic summation is performed over all possible permutations of components of vector ${\bf T^\prime}$, 
denoted as $\kappa$, and $\bf s\sqrt{T_\kappa}$$=$$[s_1\sqrt{t_{\kappa(1):n}}, 
s_2\sqrt{t_{\kappa(2):n}}, \dots, s_n\sqrt{t_{\kappa(n):n}}]^\dagger$ is a vector of inverse branches 
of transformations $t_i$$=$$x_i^2$ and ordering {\bf G}. The expression in the second line of Eq. (\ref{eq:orderStatPDF}) 
has $2n+n(n-1)/2$ unknown parameters, including $n$ components of the vector of the mean 
(${\bf \mu}$$=$$[\mu_1,\mu_2,\dots,\mu_n]^\dagger$), and $(n^2+n)/2$ entries of the symmetric covariance 
matrix (${\bf \Sigma}$$=$$[\sigma_{ij}]_{i,j=1}^n$). One can use any estimation method to determine these 
parameters. Given the statistics of $x_i$ ($\mu_i$ and $\sigma_{ij}$), we can 
obtain the pdf ($p_i(t)$), the mean ($\mu_i^T$), the variance ($\sigma_i^T$), the skewness of 
distribution ($\gamma_i^T$) for $t_i$ ($i$$=$$1,2$), and estimate pair-wise correlations ($\rho_{ij}^T$) 
using the analysis in the Appendix, which treats the case of $i,j$$=$$1,2$.


\section{FORCE-CLAMP MEASUREMENTS {\em in silico}}

\subsection{Computer model of dimer $WW$$-$$WW$}

We used the $C_\alpha$-based Self Organized Polymer (SOP) model of the polypeptide chain \cite{SOP0} 
to describe the dimer $WW$$-$$WW$ formed by the all-$\beta$-sheet $WW$ domains (Fig.~1). The $WW$ domain 
has $34$ amino acid residues (Protein Data Bank (PDB) entry 1PIN \cite{WW1}). The mechanical unraveling 
of $WW$ is described by the single-step kinetics of unfolding, $F \rightarrow U$, from the folded state 
$F$ to the unfolded state $U$. The dimer $WW$$-$$WW$ was constructed by connecting the N- and C-termini 
of the adjacent $WW$ domain using flexible linkers of two and four neutral residues (Fig.~1). 
Each residue in $WW$$-$$WW$ was represented by its $C_\alpha$-atom with the $C_\alpha$$-$$C_\alpha$ 
covalent bond distance of $a$$=$$3.8 \AA$ (peptide bond length). 

The molecular potential energy of a protein conformation, specified in terms of the residue coordinates 
$\{{\bf r}_i\}$, $i$$=$$1,2,\dots,M$, is given by
\begin{eqnarray}\label{eq:sop}
V_{MOL} & = & V_{FENE}+V^{ATT}_{NB}+V^{REP}_{NB}\nonumber\\
        & = &-\sum_{i=1}^{M-1}{\frac{k}{2}R_0^2\log{\left(1-\frac{\left(r_{i,i+1}-r^0_{i,i+1}\right)^2}{R_0^2}\right)}}\nonumber\\
        & + & \sum_{i=1}^{M-3}\sum_{j=i+3}^{M}\varepsilon_h\left[\left(\frac{r^{0}_{ij}}{r_{ij}}\right)^{12}-2
              \left(\frac{r^{0}_{ij}}{r_{ij}}\right)^{6}\right]\Delta_{ij}\nonumber\\
        & + & \sum_{i=1}^{M-2}\varepsilon_l\left(\frac{\sigma}{r_{i,i+1}}\right)^{6} + \sum_{i=1}^{M-3}\sum_{j=i+3}^{M}
              \varepsilon_l\left(\frac{r^{0}_{ij}}{r_{ij}}\right)^{6}(1-\Delta_{ij})
\end{eqnarray}
where the distance between any two interacting residues $i$ and $i$$+$$1$ is $r_{i,i+1}$, whereas 
$r^0_{i,i+1}$ is its value in the native structure. The first term in Eq. (\ref{eq:sop}) is the 
FENE potential, which describes the chain connectivity; $R_0$$=$$2 \AA$ is the tolerance in the 
change of a covalent bond ($k$$=$$1.4 N/m$). The second term is the Lennard-Jones potential 
($V_{NB}^{ATT}$), which accounts for the native interactions. We assumed that if the non-covalently 
linked residues $i$ and $j$ ($|i-j|$$>$$2$) are within the cutoff distance in the native state $r_C$$=$$8.0 \AA$, 
then $\Delta_{ij}$$=$$1$ and zero otherwise. We used a uniform value of $\varepsilon_h$$=$$1.5 kcal/mol$, 
which specifies the strength of the non-bonded interactions. All the non-native interactions were 
treated as repulsive ($V_{NB}^{REP}$). An additional constraint was imposed on the bond angle between 
residues $i$, $i$$+$$1$, and $i$$+$$2$ by including the repulsive potential with parameters 
$\varepsilon_l$$=$$1 kcal/mol$ and $\sigma$$=$$3.8 \AA$, which quantify the strength and the range of 
repulsion. To ensure the self-avoidance of the protein chain, we set $\sigma$$=$$3.8 \AA$.


\subsection{Simulations of forced unfolding of dimer $WW$$-$$WW$}

The unfolding dynamics were obtained by integrating the Langevin equations for each particle position 
${\bf r}_i$ in the over-damped limit, $\eta d{\bf r}_i/dt$$=$$-\partial V/\partial {\bf r}_i$$+$${\bf g}_i(t)$. 
Here, $V$$=$$V_{MOL}$$-$${\bf f}$${\bf Y}$ is the total potential energy, in which the first 
term ($V_{MOL}$) is the molecular contribution (see Eq. (\ref{eq:sop})) and the second term represents 
the influence of applied force on the molecular extension ${\bf Y}$. Also, ${\bf g}­(t)$ is 
the Gaussian distributed random force and $\eta$ is the friction coefficient. In each simulation run, 
the N-terminal $C_\alpha$-atom of the first domain ($WW_1$) was constrained and a constant force 
${\bf f}$$=$$f$${\bf n}$ was applied to the C-terminal $C_\alpha$-atom of the second domain ($WW_2$) 
in the direction ${\bf n}$ of the end-to-end vector of the dimer ${\bf Y}$ (see Fig.~1). 
The Langevin equations were propagated with the time step $\Delta t$$=$$0.08$$\tau_H$$=$$20 ps$, where 
$\tau_H$$=$$\zeta$$\varepsilon_h$$\tau_L/k_BT$. Here, $\tau_L$$=$$(ma^2/\varepsilon_h)^{1/2}$$=$$3 ps$, 
$\zeta$$=$$50$ is the dimensionless friction constant for a residue in water ($\eta$$=$$\zeta m/\tau_L$), 
and $m$$\approx$$3$$\times$$10^{-22} g$ is the residue mass \cite{LD1}. Pulling simulations were 
carried out at room temperature using the bulk water viscosity, which corresponds to the friction 
coefficient $\eta$$=$$7.0$$\times$$10^5 pN ps/nm$. We utilized the GPU-based acceleration to generate
the statistically representative sets of the unfolding time data \cite{GPU1,GPU2}. For each value of 
constant force $f$$=$$80$, $100$, $120$, $140$ and $160 pN$, we generated two sets of 
trajectories with $1,000$ trajectories in each set: one set for dimer of $WW$ domains connected by 
the linker of two neutral residues, and the other for the four-residue linker. The unfolding time for 
each domain (parent statistics) was defined as the first time at which the end-to-end distance of the 
domain exceeded $90\%$ ($\sim$$11.25 nm$) of its contour length $L$$=$$33$$a$$\approx$$12.5 nm$. 

Representative trajectories of the total end-to-end distance for the dimer $WW$$-$$WW$ obtained 
at $f$$=$$100 pN$ and $f$$=$$160 pN$ are compared in Fig.~2, which also shows graphically the 
definition of the first unfolding time $t_{1:2}$ and the second unfolding time $t_{2:2}$. We observe 
a typical ``unfolding staircase'', i.e. a series of sudden step-wise increases in the 
end-to-end distance of $WW$$-$$WW$ as a function of time. These mark the consecutive unfolding transitions 
in $WW$ domains, which occur at the first unfolding time $t_{1:2}$ ($1$-st unfolding event) and at 
the second unfolding time $t_{2:2}$ ($2$-nd unfolding event). The unfolding transitions are more 
discrete at a lower force $f$$=$$100 pN$ (Fig.~2a) but more continuous at a higher force $f$$=$$160 pN$ 
(Fig.~2b). We remind that in experiment (but not in simulations), there is no way of knowing 
which domain has unfolded at any given time, and the time-ordered data ($t_{r:n}$) are the only observable 
quantities.


\section{TOPOLOGICAL COUPLING AND MECHANICAL SYMMETRY BREAKING}

To provide a basis for the Order statistics inference, we performed a direct statistical analysis of 
the simulaiton output for $WW$$-$$WW$ generated at $f$$=$$80$, $100$, $120$, $140$, and $160 pN$ 
(parent data). The data sets contain $Q$$=$$1,000$ unfolding times for each force value and for each 
linker length. First, we constructed the histogram-based estimates of the unfolding times for the 
first domain ($WW_1$) and second domain ($WW_2$), which are compared in Fig.~3 (bin size was chosen 
using the Freedman-Diaconis rule \cite{NonParam}). We calculated the average unfolding times $\mu_1^T$ 
and $\mu_2^T$ ($\mu_i^T$$=$$1/Q\sum_{j=1}^{Q}t_{ij}$, $i$$=$$1,2$), the standard deviations $\sigma_1^T$ 
and $\sigma_2^T$ ($\sigma_i$$=$$(E(t_i^2)-(\mu_i^T)^2)^{1/2}$, where $E(t_i^2)$ is the second moment 
of $t_i$, $i$$=$$1,2$), and the values of the skewness of distributions $\gamma_1^T$ and $\gamma_2^T$, 
and the Pearson correlation coefficient $\rho^{T}$$=$$\rho^{T}_{12}$. The skewness of the distributions 
was calculated using the formula $\gamma_i^T$$=$$(E(t_i^3)-3\mu_i^T(\sigma_i^T)^2-(\mu_i^T)^3)/(\sigma^T_i)^3$, 
where $E(t_i^3)$ is the third moment of $t_i$ ($i$$=$$1,2$). Pearson correlation coefficient was calculated 
as $\rho^{T}$$=$$(E(t_1 t_2)-\mu_1^T\mu_2^T)/\sigma_1^T\sigma_2^T$, where $E(t_1 t_2)$ is the first 
moment of the product $t_1 t_2$. These statistical measures are compared in Table~I and II for the case 
of the linker of two and four residues, respectively. 

The parent distributions of unfolding times are skewed, broadly distributed, and exponential-like
at low forces ($f$$=$$80 pN$) and become more narrowly distributed, and Gaussian-like 
at high forces ($f$$=$$160 pN$; see Fig.~3). These changes become manifest when comparing 
the values of $\mu_i^T$, $\sigma_i^T$, and $\gamma_i^T$, $i$$=$$1,2$ (see Tables~I and II).
Here, $\mu_1^T$ and $\mu_2^T$ decrease as $f$ is increased, because the applied force destabilizes the 
native state of $WW$, and, hence, decreases their lifetimes. At low forces, dynamic fluctuations 
are not suppressed and the unfolding transitions are more variable (stochastic), which is reflected in 
the width of the distributions ($\sigma_i^T$). When $f$ is increased, fluctuations become less important 
and the unfolding events become increasingly more deterministic (less stochastic), which results in 
the decrease of $\sigma_1^T$ and $\sigma_2^T$. 

Interestingly, we found that the unfolding times $t_1$ for the first domain ($WW_1$) and $t_2$ for the
second domain ($WW_2$) are uncorrelated at low forces, but become correlated (dependent) at a high force 
($f$$=$$160 pN$). This is reflected in the values of the correlation coefficient $\rho^{T}$ (Table~I and II). 
Hence, our results show that tension couples otherwise non-interacting protein domains forming a multi-domain 
protein. Because this result could have been observed in the case of physically interacting protein domains,
e.g., through a common binding interface, we termed this effect the ``topological coupling'' to stress the 
importance of chain connectivity (topology). Hence, our results indicate that under large 
mechanical stress protein domains might become topologically coupled even in the absence of domain-domain 
interactions. 

Another interesting finding is that although the unfolding times for the first domain 
$WW_1$ ($t_1$) and for the second domain $WW_2$ ($t_2$) are similarly distributed 
at a low force ($f$$=$$80 pN$), these become more and more nonidenticaly distributed at higher 
forces ($f$$=$$100$$-$$160 pN$). Indeed, the values of $\mu_i^T$, $\sigma_i^T$, 
and $\gamma_i^T$ for $i$$=$$1,2$ are very similar at $f$$=$$80 pN$, yet, very different at 
$f$$=$$160 pN$ (Tables~I and II). Hence, our results indicate that increased tension breaks 
the mechanical symmetry. Indeed, identical protein domains $WW_1$ and $WW_2$ have very similar
mechanical properties at a low force, but these become distinctly different at higher forces.


\section{FROM TIME-ORDERED DATA TO PARENT DISTRIBUTIONS}

In force-clamp single-molecule experiments on multidomain proteins, experimentalists have no prior knowledge 
regarding the type of random variables measured. Our results for the dimer $WW$$-$$WW$ indicate that the 
mechanical force can topologically couple the non-interacting protein domains and can break the similarity
in their physical properties even when domains are identical. Using our classification 
of random variables, the unfolding times for identical protein domains in a multi-domain protein might form 
a set of {\em iid} random variables at low forces, {\em inid} random variables at the intermediate force level, 
and {\em dnid} random variables at high enough forces. Hence, a unified approach is needed to analyze and 
model the different types of random variables. Here, we describe the results of application of Order statistics 
inference, developed in Section II, to characterize the forced unfolding times for the dimer 
$WW$$-$$WW$ obtained for different values of applied constant force $f$$=$$80$$-$$160 pN$. The formalism 
adapted for the two-domain protein is presented in the Appendix.   

Because in a single-molecule experiment size of the data sample might be small, we picked at 
random $330$ data points ($330$ pairs $(t_1,t_2)$) out of $Q$$=$$1,000$ observations from each data set. 
Next, we ordered the data for each pair to generate ordered unfolding times as observed in experiment. 
As an example for the dimer $WW$$-$$WW$ (linker of two residues), we present the 
histogram-based estimates of the pdf's of the $1$-st unfolding time ($t_{1:2}$) and the $2$-nd unfolding time 
($t_{2:2}$), obtained at $f$$=$$80$, $120$, and $160 pN$ in Fig.~4. The histograms of the ordered 
unfolding times are markedly different in terms of their overall shape, width, and position of the maximum 
(most probable unfolding time). That the maximum for $t_{2:2}$ corresponds to longer times 
compared to the maximum for $t_{1:2}$ is not unexpected, since, by construction, $t_{1:2}$$<$$t_{2:2}$. 
A surprising element is that the histograms of $t_{2:2}$ have longer tais, which means that fluctuations 
play a more important role in the unfolding transitions that occur later in time. According to 
our formalism, the time-ordered data correspond to vectors of order statistics ${\bf T'}$$=$$[t_{1:2},t_{2:2}]^{\dagger}$ 
described by the joint pdf $p_{{\bf T}'}(t_{1:2},t_{2:2})$ (see Eq. (\ref{eq:orderStat2dPDF})). 

The model for two-domain protein $WW$$-$$WW$ has five unknown parameters: the average values 
$\mu_1$ and $\mu_2$, the standard deviations $\sigma_{11}$ and $\sigma_{22}$, and the covariance $\sigma_{12}$. 
These describe the statistics of vector ${\bf X}$$=$$[x_1,x_2]^{\dagger}$. Once determined, they can be 
used to describe the parent statistics of vector ${\bf T}$$=$$[t_1,t_2]^{\dagger}$. We employed 
the Maximum Likelihood Estimation (MLE) to obtain the vector of parameters, 
${\bf \theta}$$=$$(\mu_1,\mu_2,\sigma_{11},\sigma_{22},\sigma_{12})$. The likelihood function for the pdf 
given by Eq. (\ref{eq:orderStat2dPDF}) is 
$L(\theta)$$=$$\prod_{j=1}^{330}p_{\bf T^\prime }(\{t_{1:n},t_{2:n}\}_j|{\bf \theta})$, but in the model 
calculations we used the log-likelihood function, 
$\log [L(\theta)]$$=$$\sum_{j=1}^{330}\log p_{\bf T^\prime}(\{t_{1:n},t_{2:n}\}_j|{\bf \theta})$, to 
obtain the values of $\mu_1$, $\mu_2$, $\sigma_{11}$, $\sigma_{22}$, and $\sigma_{12}$. 
These quantities and Eqs. (\ref{eq:mean}), (\ref{eq:std}), (\ref{eq:skew}), and (\ref{eq:correl}) were 
then used to obtain the parent statistics: the average quantities $\mu_1^T$ and $\mu_2^T$, the standard 
deviations $\sigma_1^T$ and $\sigma_2^T$, the skewness coefficients $\gamma_1^T$ and $\gamma_2^T$, and the 
correlation coefficient $\rho^T$ (see Appendix). Finally, the closed-form expressions for the 
parent pdf's of unfolding times $p_1(t)$ and $p_2(t)$ were obtained by integrating out 
$t_2$ and $t_1$, respectively, in Eq. (\ref{eq:parent2dPDFsimpl}) for the joint parent pdf:
\begin{equation}\label{eq:pdfParenti}
p_i(t)=\frac{1}{\sigma_{ii}\sqrt{2\pi}}\cdot\frac{1}{2\sqrt{t}}\Bigl[\exp(-\frac{(\sqrt{t} - \mu_i)^2}
{2\sigma_{ii}^2}) + \exp(-\frac{(\sqrt{t} + \mu_i)^2}{2\sigma_{ii}^2}\Bigr]\text{, }i=1,2
\end{equation}
Statistical measures of the parent unfolding times (parent statistics of vector ${\bf T}$$=$$[t_1,t_2]^{\dagger}$), 
$\mu_1^T$, $\mu_2^T$, $\sigma_1^T$, $\sigma_2^T$, and $\rho^T$, obtained by applying Order statistics 
inference to the order-statistics data, are compared with the same quantities, obtained from a direct 
statistical analysis of the parent data, in Tables~I and II. The theoretical curves of the 
parent distributions $p_i(t)$ ($i=1,2$), generated by substituting the values of $\mu_1$, 
$\mu_2$, $\sigma_{11}$ and $\sigma_{22}$ (statistics of vector ${\bf X}$$=$$[x_1,x_2]^{\dagger}$) 
into Eq.(\ref{eq:pdfParenti}), are overlaid in Fig.~3 with the histograms of the parent unfolding times. 

We witness a very good agreement between the histogram-based estimates and theoretical curves of the pdf's 
of the parent unfolding times for both domains $WW_1$ and $WW_2$ and for all force values $f$$=$$80$, 
$120$, and $160 pN$ in terms of the overall shape and position of the maximum (Fig.~3). In agreement with 
simulations, our theory predicts that at high forces the second domain ($WW_2$), to which the force is applied 
(Fig.~1), unravels on a faster timescale compared to the first domain ($WW_1$). Hence, our theory captures 
the growing inequality of the time distributions for $WW_1$ and $WW_2$ at larger forces reflecting
the mechanical symmetry breaking. Comparing the values of $\mu_i^T$, $\sigma_i^T$, $\gamma_i^T$, and $\rho^T$, 
obtained using Order statistics inference, with the ``true'' values of these quantities from a direct 
statistical analysis, we see that the agreement is nearly quantitative for $\mu_i^T$ and $\sigma_i^T$ and 
very good for $\gamma_i^T$ at low forces ($f$$=$$80 pN$). At larger forces ($>$$100 pN$), 
the agreement for $\mu_i^T$ and $\sigma_i^T$ is very good, yet, the agreement for $\gamma_i^T$ is more 
qualitative rather than quantitative. Importantly, Order statistics inference correctly captured the 
mutual independence of unfolding times ($t_1$ and $t_2$) at low forces, which is reflected 
in small values of Pearson correlation coefficient $\rho^T$ for $f$$<$$140 pN$. Order statistics 
inference correctly predicts the emergence of topological interactions at higher forces (i.e. for $f$$=$$160 pN$), 
for which the values of $\rho^T$ are in the $0.1$$-$$0.3$-range (Tables~I and II).


\section{DISCUSSION}

Single-molecule force spectroscopy has enabled researchers to uncover the mechanism of adaptation 
of protein structures to mechanical loads \cite{AFMRief,RiboswitchBlock}. These experiments are now routinly 
used to study proteins and other biomolecules beyond the ensemble average picture and to map 
the entire distributions of the relevant molecular characteristics \cite{Mukamel1,Mukamel2}. Yet, existing 
theoretical approaches for analyzing and modeling the experimental results lag behind. This calls 
for the development of next generation theoretical methods, which take into account both the 
complexity of the problem and the nature and statistics of experimental observables.  

In the previos studies, we have demonstrated that the unfolding times observed in the constant force 
(force-clamp) measurements on a multidomain protein $D_1$$-$$D_2$$-$$\ldots$$-$$D_n$ comprize a set of 
the $r$-th order statistics, $t_{r:n}$ ($r$$=$$1,2,\ldots,n$) \cite{OS1}, and that the unfolding forces 
observed in the constant velocity (force-ramp) measurements form a set of the first order statistic 
$t_{1:n}$ in a sample of decreasing size ($n,n-1,\ldots,1$) \cite{OS3}. We have also developed 
rigorous statistical tests for classification of random variables measured in these experiments
\cite{OS3}. Here, we developed an Order statistical theory to describe coupled proteins forming a multimeric 
protein or proteins forming the tertiary structure within the same multidomain protein subject to the mechanical 
stress. We focused on interesting new effects of tension-induced mechanical symmetry breaking 
and topological coupling in multi-domain proteins formed by the physically non-interacting protein domains, 
using an example of the two-domain protein $WW$$-$$WW$. However, our theory can also be used to characterize 
the physico-chemical properties of proteins with strong domain-domain interactions, such as fibrin fibers, 
microtubules, and actin filaments mentioned in the introduction. Strong interactions will translate into 
large values of the off-diagonal elements of the covariance matrix (Pearson correlation coefficient).

Our approach is based on Order statistics inference, which proved to be successful at solving the inverse 
problems. The approach can be used to accurately analyze, interpret, and model the results of protein 
forced unfolding measuremens available from the constant force assays on multimeric and multidomain proteins. 
One of the main results of this paper is that Order statistics inference enables one to analyze 
and model the forced unfolding times for a multi-domain protein formed by noninteracting identical domains 
({\em iid} case) and non-identical domains ({\em inid} case), and by interacting identical domains ({\em did} 
case) and non-identical domains ({\em dnid} case). With little effort, the formalism can be extended to 
analyze the results of constant velocity measurements. We applied our approach to analyze the results of 
protein forced unfolding {\em in silico} for the dimer $WW$$-$$WW$ of the all-$\beta$-sheet $WW$ domains 
\cite{WW2,WW3}. In simulations, one can access ``the parent data'' and ``the time-ordered data''. This 
was used to compare directly the various statistical measures - the distributions of unfolding times, the 
average unfolding times, the standard deviations, and the skeweness of the distributions, obtained from the 
statistical analysis of the parent unfolding times for domains $WW_1$ and $WW_2$, and the same measures, 
obtaineed by applying Order statistics inference to the time-ordered data. A very good agreement obtained 
between the statistics of unfolding times from pulling simulations and from the theoretical inference validates 
our theory. The presense of correlations is reflected in the inequality of the joint distribution to the 
product of the marginals, i.e. $p_{\bf T}(t_1,t_2)$$\neq$$p_1(t)$$p_2(t)$ 
\cite{Mukamel1,Mukamel2}. In the bivariate case for just two domains, dynamic correlations between the 
parent unfolding times $t_1$ and $t_2$ are contained in the off-diagonal matrix element $\sigma_{12}^T$ of 
the covariance matrix ${\bf \Sigma}$ for vector ${\bf X}$$=$$[x_1,x_2]^{\dagger}$. Order statistics inference 
correctly detected the absence of correlations at a small force ($f$$=$$80$, $100$ and $120 pN$), 
and the presence of correlations at a large force ($f$$=$$160 pN$).     

We found the mechanical symmetry breaking and topological interactions observed in a multi-domain
protein subject to tension. The growing asymmetry in the mechanical properties of otherwise identical 
protein domains comprising a multi-domain protein, can be understood by considering an example of $WW$$-$$WW$. 
When the pulling force is applied, it takes time $\tau_f$ for tension to propagate from the tagged 
residue in domain $WW_2$ to the other domain $WW_1$. Hence, domain $WW_2$ is subjected to 
the mechanical force for a longer time than domain $WW_1$ (Fig.~1). When force is large enough so that 
the average unfolding time $\mu_f$ becomes comparable with $\tau_f$, tension propagation prolongs the 
lifetime of the more distal domain $WW_1$, i.e. $\mu_1(f)$$\approx$$\mu_f$$+$$\tau_f$ versus $\mu_2(f)$$\approx$$\mu_f$. 
This is reflected in the inequality of the distributions of unfolding times for domains 
$WW_1$ and $WW_2$ (Fig.~3). But this same result would have been observed for a different two-domain 
protein, say $D_1$$-$$D_2$, formed by the non-identical domains, e.g., the mechanically stronger domain 
$D_1$ and mechanically weaker domain $D_2$, characterized by the differently distributed unfolding times.

The topological coupling can be understood using a concept of correlation length. In the 
folded state, overall bending flexibility is determined by the mobility at the domain-domain interface. 
Under low tension, the average inter-domain angle $\theta$ should show large deviations from the 
$180^{\circ}$-angle, $\Delta \theta$, and the correlation length $l_c$ is short. Under high 
tension, $\Delta \theta$ decreases and $l_c$ increases. This same result would have been observed for 
the domains that interact, e.g., through their inter-domain interface. Domain interactions would have 
decreased the mobility at the domain-domain interface, which, in turn, would have resulted in smaller 
$\Delta \theta$ and longer $l_c$. The averaged squared deviations of the inter-domain angle can be linked 
to the correlation length as $\langle \Delta\theta^2 \rangle$$=$$2d/l_c$, where $d$ is the inter-domain 
distance. We estimated $l_c$ using the results of pulling simulations for $WW$$-$$WW$ (linker of two 
residues). For $f$$=$$100 pN$, $d$$\approx$$1.2 nm$ and $|\Delta\theta|$$\approx$$30^{\circ}$, and 
$l_c$$\approx$$1.8 nm$ is shorter than the average size of the $WW$ domain at equilibrium ($\approx$$3$$-$$4 nm$). 
For $f$$=$$160 pN$, $d$$\approx$$2.8 nm$ and $|\Delta\theta|$$\approx$$15^{\circ}$. This results in 
the three-fold increase in $l_c$$\approx$$5.2 nm$, which now exceeds the dimension of 
$WW$ domain. Hence, under high tension $WW$ domains unravel in a concerted fasion, which 
is reflected in the dependence of the unfolding times $t_1$ and $t_2$ at large forces (Tables~I and II).   

To provide the reader with yet another glimpse of the hidden complexity underlying a seemingly trivial 
problem of unfolding of a multi-domain protein, we performed pulling simulations for the trimer 
$WW$$-$$WW$$-$$WW$, in which the force was applied to the third domain ($WW_3$). The histograms of 
unfolding times for the first domain $WW_1$ ($t_1$), second domain $WW_2$ ($t_2$), and third domain 
$WW_3$ ($t_3$) are compared in Fig.~5. Numerical values of the correlation coefficients for pair-wise 
correlations $\rho^T_{ij}$, $i$$\neq$$j$$=$$1,2,3$, of the unfolding times are summarized in Table~III. 
We see that the distributions of unfolding times are nearly identical and unimodal, reflecting the 
mechanical symmetry, at a low $100 pN$-force; yet, the time distributions become increasingly 
more different with the increasing tension. Indeed, the distributions of unfolding times for $WW_1$ 
and $WW_2$, but not for $WW_2$, develop the second mode at $f$$=$$140 pN$. The parent unfolding times 
for $WW_1$ ($t_1$) and $WW_2$ ($t_2$), but not for $WW_1$ and $WW_3$ and not for $WW_2$ and $WW_3$, 
develop correlations, which tend to grow with force (Table~III).


\section{CONCLUSION}

Intramolecular and intermolecular interactions involving proteins are at the core of virtually 
every biological process. Here, we developed and tested a new Order statistics approach for detecting and 
describing biomolecular interactions. Order statistics offers a new tool kit for accurate interpretation 
and modeling of experimental data on multidomain proteins, multimeric proteins, and engineered polyproteins 
available from single-molecule force spectroscopy. Looking into the future, we anticipate that Order statistics 
based calculus of probability will play an important role when single-molecule techniques will expand into 
new avenues of research such as folding of multi-domain proteins \cite{WangJACS}, protein folding 
in cellular environment, and nanomechanics of protein assemblies (protein fibers, microtubules, actin filaments, 
viruses, etc.) \cite{Fg1,Fg2}. Also, life processes in living cells are coordinated both spatially and temporally. 
In this respect, Order statistics builds in causality into a theoretical description in a natural way. 

Mechanical symmetry breaking and topological coupling might be related to the biological utility of proteins. 
The structural design of multidomain proteins - number of domains, linker length, etc., carves out a 
larger ``parameter space'' for the physical characteristics  such as mechanical 
strength, tolerance to fluctuations, correlation length, all of which are force-dependent. As we showed in 
the paper, when tension is high enough, this design permits the ``mechanical differentiation'' of protein 
domains, which become, in some sense, ``functional isoforms'' of the same structural unit. Protein systems 
may have evolved to select certain modular architectures with the optimal physico-chemical properties for 
efficient mechanical integration of forces. Also, the mechanical stress promotes coupling between structural 
elements, which transforms a mechanically non-cooperative system into a highly cooperative one. This might 
be a mechanism of information transfer over long distances to coordinate cell processes over the relevant 
lengthscales and timescales \cite{MTDWang}.


\noindent 
{\bf Acknowledgments:} This work was supported by the American Heart Association (Grant 09SDG2460023) 
and by the Russian Ministry of Education and Science (Grant 14.A18.21.1239)


\appendix

\section*{APPENDIX: ORDER STATISTICS INFERENCE FOR TWO-DOMAIN CASE}

\setcounter{equation}{0}
\renewcommand{\thesection}{A}

In the case of a two-domain protein $D_1$$-$$D_2$, the unfolding time data can be represented by vectors 
${\bf T^\prime}$$=$$[t_{1:2},t_{2:2}]^\dagger$ containing the first and second order statistic. These data 
can be used to gather information about parent vectors ${\bf T}$$=$$[t_1,t_2]^\dagger$ for the first domain 
($D_1$) and second domain ($D_2$). For each vector ${\bf T}$ we define vector ${\bf X}^2$$=$$[x_1^2,x_2^2]^\dagger$, 
in which random variables $x_1$ and $x_2$ are sampled from the Gaussian distribution (see Eq. (\ref{eq:xPDF})). 
Statistics of $x_1$ and $x_2$ are described by the vector of the average ${\bf \mu}$$=$$[\mu_1,\mu_2]^\dagger$ 
and the covariance matrix ${\bf \Sigma}$$=$$\Bigl[\begin{smallmatrix}\sigma_{11}^2 & \sigma_{12}\\\sigma_{21} & \sigma_{22}^{2}
\end{smallmatrix}\Bigr]$, where $\sigma_{11}^2$ and $\sigma_{22}^2$ are the variances of $x_1$ and $x_2$, 
respectively, and $\sigma_{12}$$=$$\sigma_{21}$ is the covariance. Eq. (\ref{eq:parentPDF}) 
for the joint pdf of the parent data ${\bf T}$$=$$[t_1,t_2]^\dagger$ becomes
\begin{equation}\label{eq:parent2dPDF}
p_{{\bf T}}(t_{1}, t_{2}) = \frac{1}{2\pi\sqrt{|{\bf \Sigma}|}} \cdot \frac{1}{2^2\sqrt{t_1 \cdot t_2}} \cdot 
\sum_{{\bf s}}{e^{-\frac{1}{2}({\bf s}{\bf \sqrt{T}} - {\bf \mu})^{\dagger}{\bf \Sigma}^{-1}
({\bf s}{\bf \sqrt{T}} - {\bf \mu})}}
\end{equation}
where $|{\bf \Sigma}|$$=$$\sigma_{11}^2\sigma_{22}^2$$-$$\sigma_{12}^{2}$ is the determinant and
${\bf \Sigma}^{-1}$$=$$\frac{1}{|{\bf \Sigma}|}\Bigl[\begin{smallmatrix}\sigma_{22}^2 & -\sigma_{12} 
\\-\sigma_{12} & \sigma_{11}^{2} \end{smallmatrix}\Bigr]$ is the inverse of the covariance matrix 
${\bf \Sigma}$. In Eq. (\ref{eq:parent2dPDF}), ${\bf s}=[s_1, s_2]^\dagger=\left\{[1,1]^\dagger, [1, -1]^\dagger, 
[-1, 1]^\dagger, [-1, -1]^\dagger\right\}$ are vectors of all possible signs of $\sqrt{t_1}$ and $\sqrt{t_2}$
and ${\bf s\sqrt{T}}$$=$$[s_1\sqrt{t_1}, s_2\sqrt{t_2}]^{\dagger}$ is the vector of inverse values, 
which correspond to all possible branches of the inverse transformation $[t_1,t_2]$$=$$[x_1^2,x_2^2]$. 
Performing vector multiplication in the exponent of the exponential function forming the sum in 
Eq. (\ref{eq:parent2dPDF}), we obtain:
\begin{eqnarray}
({\bf s}{\bf \sqrt{T}} - {\bf \mu})^{\dagger} & {\bf \Sigma}^{-1} & 
({\bf s}{\bf \sqrt{T}} - {\bf \mu})\nonumber\\
& = & \frac{(s_1\sqrt{t_1} - \mu_1)^2\sigma_{22}^2 + (s_2\sqrt{t_2} - \mu_2)^2\sigma_{11}^2 -2
(s_1\sqrt{t_1} - \mu_1)(s_2\sqrt{t_2} - \mu_2)\sigma_{12}}{\sigma_{11}^2\sigma_{22}^2 - \sigma_{12}^{2}}
\end{eqnarray}
which, depending on the choice of ${\bf s}$$=$$[s_1, s_2]^\dagger$, takes the following form:
\begin{equation}
k_1(t_1,t_2)=\frac{(\sqrt{t_1} - \mu_1)^2\sigma_{22}^2 + (\sqrt{t_2} - \mu_2)^2\sigma_{11}^2 - 
2(\sqrt{t_1} - \mu_1)(\sqrt{t_2} - \mu_2)\sigma_{12}}{2(\sigma_{12}^2 - \sigma_{11}^2\sigma_{22}^2)}
\text{, } {\bf s} = [1, 1]^\dagger\nonumber\\
\end{equation}
\begin{equation}
k_2(t_1,t_2) = \frac{(\sqrt{t_1} - \mu_1)^2\sigma_{22}^2 + (\sqrt{t_2} + \mu_2)^2\sigma_{11}^2 + 
2(\sqrt{t_1} - \mu_1)(\sqrt{t_2} + \mu_2)\sigma_{12}}{2(\sigma_{12}^2 - \sigma_{11}^2\sigma_{22}^2)}
\text{, } {\bf s} = [1, -1]^\dagger\nonumber\\
\end{equation}
\begin{equation}
k_3(t_1,t_2) = \frac{(\sqrt{t_1} + \mu_1)^2\sigma_{22}^2 + (\sqrt{t_2} - \mu_2)^2\sigma_{11}^2 + 
2(\sqrt{t_1} + \mu_1)(\sqrt{t_2} - \mu_2)\sigma_{12}}{2(\sigma_{12}^2 - \sigma_{11}^2\sigma_{22}^2)}
\text{, } {\bf s} = [-1, 1]^\dagger\nonumber\\
\end{equation}
\begin{equation}
k_4(t_1,t_2) = \frac{(\sqrt{t_1} + \mu_1)^2\sigma_{22}^2 + (\sqrt{t_2} + \mu_2)^2\sigma_{11}^2 - 
2(\sqrt{t_1} + \mu_1)(\sqrt{t_2} + \mu_2)\sigma_{12}}{2(\sigma_{12}^2 - \sigma_{11}^2\sigma_{22}^2)}
\text{, } {\bf s} = [-1, -1]^\dagger\nonumber\\
\end{equation}
Substituting these expressions for $k_j$$=$$k_j(t_1,t_2)$ ($j$$=$$1,2,3,4$) and the expression 
for $|{\bf \Sigma}|$ in Eq. (\ref{eq:parent2dPDF}), we obtain the closed form expression for the 
joint pdf of the parent data:
\begin{equation}\label{eq:parent2dPDFsimpl}
p_{{\bf T}}(t_{1}, t_{2}) = \frac{1}{2\pi\sqrt{\sigma_{11}^2\sigma_{22}^2-\sigma_{12}^{2}}} 
\cdot \frac{1}{4\sqrt{t_1 \cdot t_2}} \cdot \sum_{j=1}^{4}{e^{k_{j}(t_1, t_2)}}
\end{equation}
in which the summation over ${\bf s}$ is replaced with the summation over $j$. 

Next, to obtain the closed form expression for the joint pdf of the order statistics, 
we use Eq. (\ref{eq:orderStatPDF}) and take into account possible permutations of the vector 
components of order statistics  $[t_{\kappa(1):2},t_{\kappa(2):2}]^\dagger$­. In the 2D-case, there 
are only two options, namely vector $[t_{1:2},t_{2:2}]^\dagger$­ and vector $[t_{2:2},t_{1:2}]^\dagger$­, 
which correspond to permutations $(t_1, t_2)$­ and $(t_2, t_1)$­ of the parent data. Then, the 
joint pdf of the order statistics is given by
\begin{eqnarray}\label{eq:orderStat2dPDF}
p_{{\bf T^\prime}}(t_{1:2}, t_{2:2})& = & p_{{\bf T}}(t_{1}, t_{2}) + p_{{\bf T}}(t_{2}, t_{1})\nonumber\\
& = & \frac{1}{4\sqrt{t_1 \cdot t_2}} \cdot \frac{1}{2\pi\sqrt{\sigma_{11}^2\sigma_{22}^2-\sigma_{12}^{2}}}\nonumber\\
& \times & \Bigl[\exp(\frac{(\sqrt{t_1} - \mu_1)^2\sigma_{22}^2 + (\sqrt{t_2} - \mu_2)^2\sigma_{11}^2 - 2(\sqrt{t_1} - \mu_1)
(\sqrt{t_2} - \mu_2)\sigma_{12}}{2(\sigma_{12}^2 - \sigma_{11}^2\sigma_{22}^2)})\nonumber\\
& + & \exp(\frac{(\sqrt{t_1} - \mu_1)^2\sigma_{22}^2 + (\sqrt{t_2} + \mu_2)^2\sigma_{11}^2 + 2(\sqrt{t_1} - \mu_1)
(\sqrt{t_2} + \mu_2)\sigma_{12}}{2(\sigma_{12}^2 - \sigma_{11}^2\sigma_{22}^2)})\nonumber\\
& + & \exp(\frac{(\sqrt{t_1} + \mu_1)^2\sigma_{22}^2 + (\sqrt{t_2} - \mu_2)^2\sigma_{11}^2 + 2(\sqrt{t_1} + \mu_1)
(\sqrt{t_2} - \mu_2)\sigma_{12}}{2(\sigma_{12}^2 - \sigma_{11}^2\sigma_{22}^2)})\nonumber\\
& + & \exp(\frac{(\sqrt{t_1} + \mu_1)^2\sigma_{22}^2 + (\sqrt{t_2} + \mu_2)^2\sigma_{11}^2 - 2(\sqrt{t_1} + \mu_1)
(\sqrt{t_2} + \mu_2)\sigma_{12}}{2(\sigma_{12}^2 - \sigma_{11}^2\sigma_{22}^2)})\nonumber\\
& + & \exp(\frac{(\sqrt{t_2} - \mu_1)^2\sigma_{22}^2 + (\sqrt{t_1} - \mu_2)^2\sigma_{11}^2 - 2(\sqrt{t_2} - \mu_1)
(\sqrt{t_1} - \mu_2)\sigma_{12}}{2(\sigma_{12}^2 - \sigma_{11}^2\sigma_{22}^2)})\nonumber\\
& + & \exp(\frac{(\sqrt{t_2} - \mu_1)^2\sigma_{22}^2 + (\sqrt{t_1} + \mu_2)^2\sigma_{11}^2 + 2(\sqrt{t_2} - \mu_1)
(\sqrt{t_1} + \mu_2)\sigma_{12}}{2(\sigma_{12}^2 - \sigma_{11}^2\sigma_{22}^2)})\nonumber\\
& + & \exp(\frac{(\sqrt{t_2} + \mu_1)^2\sigma_{22}^2 + (\sqrt{t_1} - \mu_2)^2\sigma_{11}^2 + 2(\sqrt{t_2} + \mu_1)
(\sqrt{t_1} - \mu_2)\sigma_{12}}{2(\sigma_{12}^2 - \sigma_{11}^2\sigma_{22}^2)})\nonumber\\
& + & \exp(\frac{(\sqrt{t_2} + \mu_1)^2\sigma_{22}^2 + (\sqrt{t_1} + \mu_2)^2\sigma_{11}^2 - 2(\sqrt{t_2} + \mu_1)
(\sqrt{t_1} + \mu_2)\sigma_{12}}{2(\sigma_{12}^2 - \sigma_{11}^2\sigma_{22}^2)}) \Bigr]
\end{eqnarray}
To obtain the average values $\mu_1$ and $\mu_2$, the standard deviations $\sigma_{11}$ and $\sigma_{22}$, 
and the covariance $\sigma_{12}$, we used the Maximum Likelihood Estimation method (Section V) and the 
joint pdf of the order statistics (see Eq. (\ref{eq:orderStat2dPDF})). These parameters correspond to 
the vector ${\bf X}$$=$$[x_1,x_2]^\dagger$, which allows us to estimate the average value ($\mu^T_i$), 
the standard deviation ($\sigma^T_i$), the skewness of the distribution ($\gamma^T_i$), and the correlation 
coefficient ($\rho^T$) for the parent data ${\bf T}$$=$$[t_1,t_2]^\dagger$. 

The theoretically estimated average unfolding times for domains $D_1$ and $D_2$ are given by 
\begin{eqnarray}\label{eq:mean}
\mu^T_i & = & E(t_i) = E(x_i^2)\nonumber\\ 
        & = & \mu_i^2 + \sigma_{ii}^2, \quad i=1,2
\end{eqnarray}
where $E(t_i)$ denotes the expected value of $t_i$ and $E(x_i^2)$ is the second moment of $x_i$. 
The standard deviations are given by
\begin{eqnarray}\label{eq:std}
\sigma^T_i & = & \sqrt{Var(t_i)} = \sqrt{Var(x_i^2)}\nonumber\\
           & = & \sqrt{E(x_i^4) - (E(x_i^2))^2}\nonumber\\
           & = & \sqrt{4\sigma_{ii}^2\mu_i^2 + 2\sigma_{ii}^4}, \quad i=1,2
\end{eqnarray}
where $Var(t_i)$$=$$Var(x_i^2)$ is the variance of the parent unfolding times $t_i$$=$$x_i^2$,
and $E(x_i^2)$ and $E(x_i^4)$ are the second and fourth moments of $x_i$, respectively ($i$$=$$1,2$). 
The skewness of the distributions can be estimated as 
\begin{eqnarray}\label{eq:skew}
\gamma^T_i & = & \frac{E(t_i^3) - 3\mu^T_i(\sigma^T_i)^2 - (\mu^T_i)^3}{(\sigma^T_i)^3}\nonumber\\
           & = & \frac{E(x_i^6) - 3\mu^T_i(\sigma^T_i)^2 - (\mu^T_i)^3}{(\sigma^T_i)^3}\nonumber\\
           & = & \frac{24\mu_i^2\sigma_{ii}^4 + 8\sigma_{ii}^6}{\sqrt{(4\sigma_{ii}^2\mu_i^2 + 2\sigma_{ii}^4)^3}}, \quad i=1,2
\end{eqnarray}
The correlation coefficient of unfolding times $t_1$ and $t_2$, $\rho^T$$=$$\rho_{ij}$ 
($i$$\neq$$j$$=$$1,2$) is defined as
\begin{eqnarray}\label{eq:correl}
\rho^T & = & \frac{E(t_1 t_2) - E(t_1)E(t_2)}{\sqrt{Var(t_1)}\sqrt{Var(t_2)}}\nonumber\\ 
       & = & \frac{E(x_{1}^2 x_{2}^2) - E(x_{1}^2)E(x_{2}^2)}{\sqrt{Var(x_{1}^2)}\sqrt{Var(x_{2}^2)}}
\end{eqnarray}
In order to calculate $E(x_{1}^2 x_{2}^2)$ - the expected value of $x_1^2$ and $x_2^2$, we rewrite
$x_2$ in the form $x_2 = \mu_2+\rho\frac{\sigma_{22}}{\sigma_{11}}(x_1 - \mu_1) + \sigma_{22}\sqrt{1 - \rho^2}R$,
where $R$ is a normal random variable, independent of $x_1$, with zero mean and unit variance, and $\rho$ 
is the correlation coefficient for $x_1$ and $x_2$, given by $\rho=\sigma_{12}/(\sigma_{11}\sigma_{22})$. 
Then, $E(x_{1}^2 x_{2}^2)$ becomes $E(x_{1}^2 x_{2}^2) = E(x_1^2(\mu_2+\rho\frac{\sigma_{22}}{\sigma_{11}}
(x_1 - \mu_1) + \sigma_{22}\sqrt{1 - \rho^2}R)^2)$. Simplifying the expression for $E(x_{1}^2 x_{2}^2)$
we obtain:
\begin{equation}\label{eq:EX1X2}
E(x_{1}^2 x_{2}^2) = A^2E(x_1^2) + \frac{\sigma_{12}^2}{\sigma_{11}^4}E(x_1^4) + 
2A\frac{\sigma_{12}}{\sigma_{11}^2}E(x_1^3) + (\sigma_{22}^2 - \frac{\sigma_{12}^2}{\sigma_{11}^2})E(x_1^2)
\end{equation}
where the coefficient $A$, the third moment $E(x_1^3)$ of $x_1$, and the fourth moment $E(x_1^4)$ of $x_1$
are given, respectively, by
\begin{eqnarray}\label{eq:AE3E4}
A & = & \mu_2 - \frac{\sigma_{12}}{\sigma_{11}^2}\mu_1,\nonumber\\ 
E(x_1^3) & = & \mu_1^3 + 3\mu_1\sigma_{11}^2,\nonumber\\ 
E(x_1^4) & = & \mu_1^4 + 6\mu_1^2\sigma_{11}^2 + 3\sigma_{11}^4
\end{eqnarray}
The correlation coefficient for the parent data can be calculated theoretically by substituting 
Eqs. (\ref{eq:AE3E4}) into Eq. (\ref{eq:EX1X2}), then substituting Eq. (\ref{eq:EX1X2}) into the second 
line in Eq. (\ref{eq:correl}), and finally using Eqs. (\ref{eq:mean}) and (\ref{eq:std}) for the average 
unfolding time $\mu_i^T$$=$$E(x_i^2)$ and the variance $\sigma_i^T$$=$$\sqrt{Var(x_i^2)}$ ($i$$=$$1,2$).




\newpage

\noindent
{\bf FIGURE CAPTIONS}

\bigskip
\noindent
{\bf Figure 1:} Schematic representation of the native structure of dimer $WW$$-$$WW$, formed by the C-terminal 
to N-terminal connected all-$\beta$-sheet $WW$ domains (PDB code: 1PIN). The first $WW$ domain, denoted as $WW_1$, 
is shown in blue, the second $WW$ domain, denoted as $WW_2$, is shown in red, and the two-residue linker 
is shown in yellow color. In pulling simulations, the constant force $f$ is applied to the C-terminus of domain 
$WW_2$ in the direction coinsiding with the end-to-end vector of dimer $WW$$-$$WW$;
the N-terminus of domain $WW_1$ is constrained. Structural analysis of unfolding trajectories revealed 
that the force unfolding transitions from the native folded state ($F$) to the unfolded state ($U$) in 
each $WW$ domain occur in a single step, $F$$\to$$U$. Shown are the two possible scenarios. In the first pathway 
(left), $WW_1$ unfolds first ($1$-st unfolding time $t_{1:2}$) and $WW_2$ unfolds second ($2$-nd unfolding time 
$t_{2:2}$). In the second pathway (right), $WW_1$ unfolds second ($t_{2:2}$) and $WW_2$ unfolds first ($t_{1:2}$).

\bigskip
\noindent
{\bf Figure 2:} The time-evolution of the end-to-end distance of the dimer $WW$$-$$WW$ (see Fig.~1), $Y$, under 
the influence of constant pulling force of $f$$=$$100 pN$ (panel $a$) and $f$$=$$160 pN$ (panel $b$). Shown in 
different color are a few representative trajectories. The unfolding transitions in $WW$$-$$WW$ are reflected in 
the stepwise increases in $Y$, which occur at the $1$-st unfolding time $t_{1:2}$ (unfolding of $WW_1$ or $WW_2$), 
and $2$-nd unfolding time $t_{2:2}$ (unfolding of $WW_1$ or $WW_2$). These transitions are magnified in {\em the insets} 
for each force value for just one simulation run.

\bigskip
\noindent
{\bf Figure 3:} The ``parent unfolding times'' $t_i$, $i$$=$$1,2$, for the dimer $WW$$-$$WW$ of 
domains $WW_1$ and $WW_2$ connected by the two-residue linker (panels $a$$-$$f$) and four-residue linker (panels 
$g$$-$$l$). Shown are the histogram-based estimates of the pdf's of unfolding times for the first domain $WW_1$ 
($t_1$, blue bars) and for the second domain $WW_2$ ($t_2$, red bars), obtained directly from the simulation 
output. These are compared with the theoretical curves of the same quantities obtained by applying Order statistics 
inference (black curves) for $f$$=$$80 pN$ (panels $a$, $b$ and $g$, $h$), $f$$=$$120 pN$ (panels $c$, 
$d$ and $i$, $j$), and $f$$=$$160 pN$ (panels $e$, $f$ and $k$, $l$).

\bigskip
\noindent
{\bf Figure 4:} The ``time-ordered unfolding times'' $t_{r:2}$, $r$$=$$1,2$ for the dimer $WW$$-$$WW$ 
(two-residue linker). The data are represented by the histogram-based estimates of the pdf's of the $1$-st unfolding 
time ($t_{1:2}$) and $2$-nd unfolding time ($t_{2:2}$) compared for $f$$=$$80 pN$ (panels $a$ and 
$b$), $f$$=$$120 pN$ (panels $c$ and $d$), and $f$$=$$160 pN$ (panels $e$ and $f$).

\bigskip
\noindent
{\bf Figure 5:} The ``parent unfolding times'' $t_i$, $i$$=$$1,2$ and $3$, for the trimer $WW$$-$$WW$$-$$WW$
of $WW$ domains connected by linkers of two residues. Compared are the histogram-based estimates of the parent 
pdf's of unfolding times for the first domain $WW_1$ ($t_1$, blue bars), second domain $WW_2$ ($t_2$, red bars), 
and third domain $WW_3$ ($t_3$, green bars), obtained directly from the simulation output generated for $f$$=$$100 pN$ 
(panels $a$, $b$, and $c$), $f$$=$$140 pN$ (panels $d$, $e$, and $f$), and $f$$=$$160 pN$ (panels $g$, $h$, and $i$).


\bigskip
\noindent
{\bf Table I:} Statistical measures of the parent unfolding times for the first domain $WW_1$ ($t_1$), 
and second domain $WW_2$ ($t_2$), connected in the dimer $WW$$-$$WW$ by the two-residue linker: the 
average unfolding times $\mu^T_1$ and $\mu^T_2$, the standard deviations $\sigma^T_1$ and $\sigma^T_2$, 
the skewness of the distributions $\gamma^T_1$ and $\gamma^T_2$, the Pearson correlation coefficient 
$\rho^T$$=$$\rho_{12}^T$, and Spearman rank correlation coefficient $s^T$$=$$s_{12}^T$. 
The estimates of these measures (except for $s^T$), obtained directly from the simulation output, are compared 
with the estimates obtained by applying Order statistics inference (shown in parentheses). 
\begin{center}
\begin{tabular}{c*{7}{c}c}
\hline \hline
\multirow{2}{*}{Force,} & \multirow{2}{*}{$\mu^T_1$,} & \multirow{2}{*}{$\mu^T_2$,} & \multirow{2}{*}{$\sigma^T_1$,} & \multirow{2}{*}{$\sigma^T_2$,} & \multirow{2}{*}{$\gamma^T_1$}  & \multirow{2}{*}{$\gamma^T_2$} & \multirow{2}{*}{$\rho^T$} & \multirow{2}{*}{$s^T$} \\
pN & ms & ms & ms & ms &  &  &  \\ 
\hline 
\multirow{2}{*}{80}   &  174.2  &  106.1  &  134.4  &  102.3  &  1.69  &  1.51  &  -0.012 & 0.006  \\
      & (189.8) & (93.5) & (143.7) & (88.5) & (1.2) & (1.55) & (-0.097) &   \\
\multirow{2}{*}{100}   &  4.13  &  2.25  &  3.60  &  2.25  &  1.41  &  1.67  &  -5$\cdot10^{-5}$ & 0.025  \\
      & (4.19) & (2.09) & (3.77) & (1.87) & (1.46) & (1.67) & (-0.067) &   \\
\multirow{2}{*}{120}   &  0.32  &  0.13  &  0.23  &  0.12  &  1.71  &  1.98  &  -0.064  & -0.049 \\
      & (0.31) & (0.13) & (0.23) & (0.08) & (1.20) & (0.98) & (-0.111) &  \\
\multirow{2}{*}{140}   &  0.093  &  0.034  &  0.031  &  0.019  &  1.21  &  2.01  &  0.014 & 0.132  \\
      & (0.095) & (0.032) & (0.024) & (0.016) & (0.39) & (0.76) & (0.174)  \\
\multirow{2}{*}{160}   &  0.058  &  0.016  &  0.010  &  0.006  &  1.18  &  1.86  &  0.131 & 0.123 \\
      & (0.057) & (0.016) & (0.009) & (0.005) & (0.24) & (0.51) & (0.151) & \\
\hline \hline
\end{tabular}
\end{center}


\newpage
\noindent
{\bf Table II:} Same quantities as in Table I but for the dimer $WW$$-$$WW$ of $WW$ domains 
connected by the linker of four residues.  
\begin{center}
\begin{tabular}{c*{7}{c}c}
\hline \hline
\multirow{2}{*}{Force,} & \multirow{2}{*}{$\mu^T_1$,} & \multirow{2}{*}{$\mu^T_2$,} & \multirow{2}{*}{$\sigma^T_1$,} & \multirow{2}{*}{$\sigma^T_2$,} & \multirow{2}{*}{$\gamma^T_1$}  & \multirow{2}{*}{$\gamma^T_2$} & \multirow{2}{*}{$\rho^T$} & \multirow{2}{*}{$s^T$} \\
pN & ms & ms & ms & ms &  &  &  \\ 
\hline 
\multirow{2}{*}{80}   &  166.0  &  112.9  &  133.5  &  99.6  &  1.74  &  1.35  &  0.028  & 0.037  \\
      & (150.0) & (103.8) & (125.4) & (94.3) & (1.34) & (1.47) & (0.077) &  \\
\multirow{2}{*}{100}   &  3.80  &  2.46  &  3.30  &  2.30  &  1.42  &  1.62  &  0.030  & 0.052  \\
      & (4.04) & (2.52) & (3.59) & (1.97) & (1.44) & (1.24) & (0.048) &  \\
\multirow{2}{*}{120}   &  0.32  &  0.15  &  0.23  &  0.12  &  1.68  &  1.73  &  -0.026 & -0.033 \\
      & (0.30) & (0.16) & (0.23) & (0.10) & (1.20) & (0.96) & (-0.004) &  \\
\multirow{2}{*}{140}   &  0.10  &  0.040  &  0.036  &  0.023  &  1.42  &  1.55  &  -0.017 & 0.069  \\
      & (0.10) & (0.039) & (0.032) & (0.021) & (0.46) & (0.83) & (-0.266) &  \\
\multirow{2}{*}{160}   &  0.065  &  0.021  &  0.011  &  0.008  &  1.19  &  1.00  &  0.259 & 0.255 \\
      & (0.064) & (0.021) & (0.011) & (0.008) & (0.25) & (0.55) & (0.285)  & \\
\hline \hline
\end{tabular}
\end{center}


\bigskip
\noindent
{\bf Table III:} The Pearson correlation coefficients, $\rho^T_{ij}=\rho_{12}^T$, $\rho_{13}^T$, 
and $\rho_{22}^T$, and the Spearman rank correlation coeficients, $s^T_{ij}=s_{12}^T$, $s_{13}^T$, 
and $s_{22}^T$, quantifying the degree of pairwise correlations (dependence) of the parent unfolding 
times $t_i$ ($i$$=$$1,2,3$) for the first domain $WW_1$ ($t_1$), second domain $WW_2$ ($t_2$), and third 
domain $WW_3$ ($t_3$). These measures are obtained by performing a statistical analysis of the 
simulation output for the trimer $WW$$-$$WW$$-$$WW$ of $WW$ domains connected by the linkers of two residues.
\begin{center}
\begin{tabular}{c*{4}{c}c}
\hline \hline
Force, pN & $\rho_{12}^T$($s_{12}^T$) & $\rho_{13}^T$($s_{13}^T$) & $\rho_{23}^T$($s_{23}^T$) \\
\hline 
100       & -0.023 (0.032)  &  0.053  (0.050) &  0.038  (0.035) \\
120       & -0.073 (-0.130) & -0.035 (-0.039) &  0.054 (-0.032) \\
140       & -0.341 (-0.295) &  0.042  (0.057) &  0.043  (0.092) \\
160       & -0.537 (-0.139) &  0.038  (0.021) & -0.016 (-0.032) \\
\hline \hline
\end{tabular}
\end{center}


\newpage
\begin{figure}[h]\label{fig:figure1}
\includegraphics[width=6.5in]{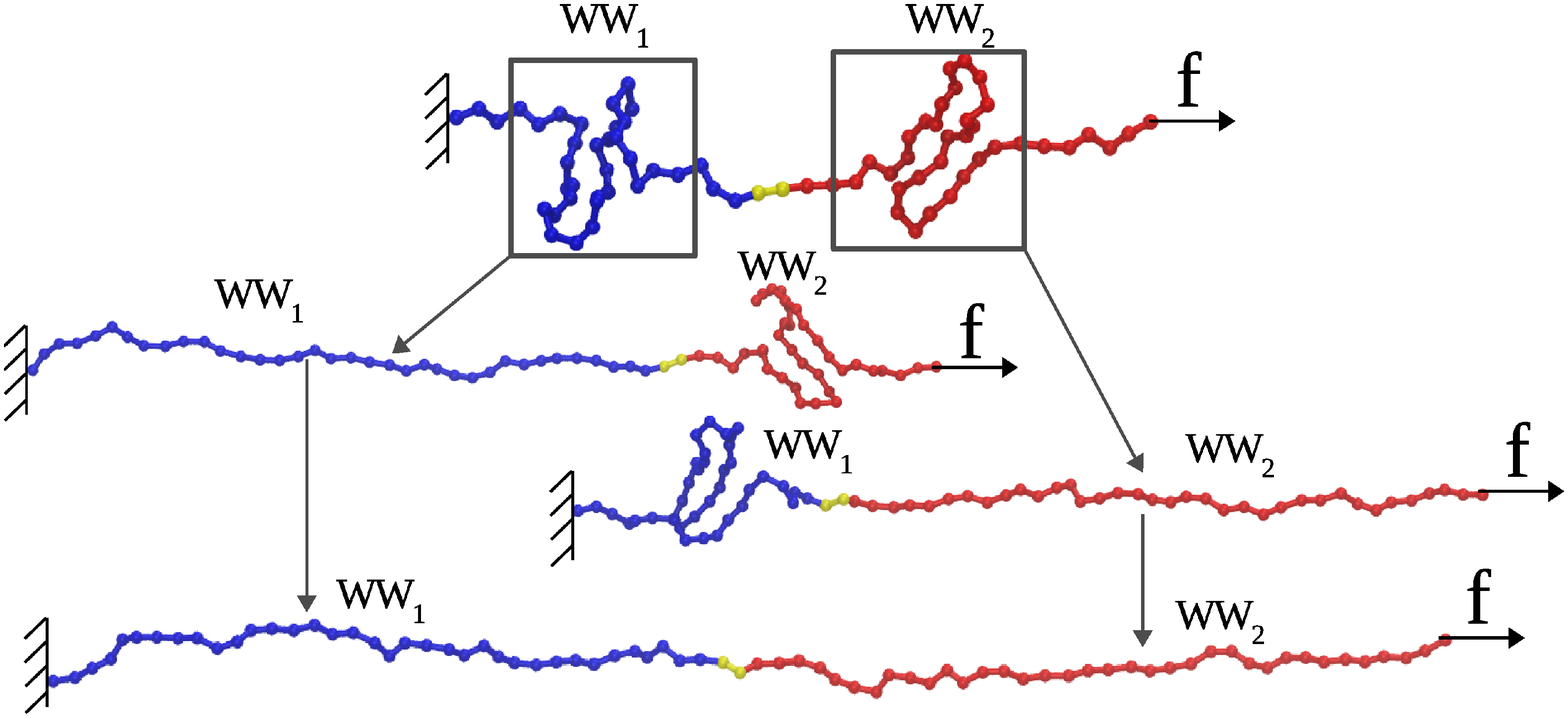}
\caption{{\large(Olga Kononova, Lee Jones, Valeri Barsegov)}}
\end{figure}


\newpage
\begin{figure}[h]\label{fig:figure2}
\includegraphics[width=6.5in]{WW_Fig2.eps}
\caption{{\large(Olga Kononova, Lee Jones, Valeri Barsegov)}}
\end{figure}


\newpage
\begin{figure}[h]\label{fig:figure3}
\includegraphics[width=4.8in]{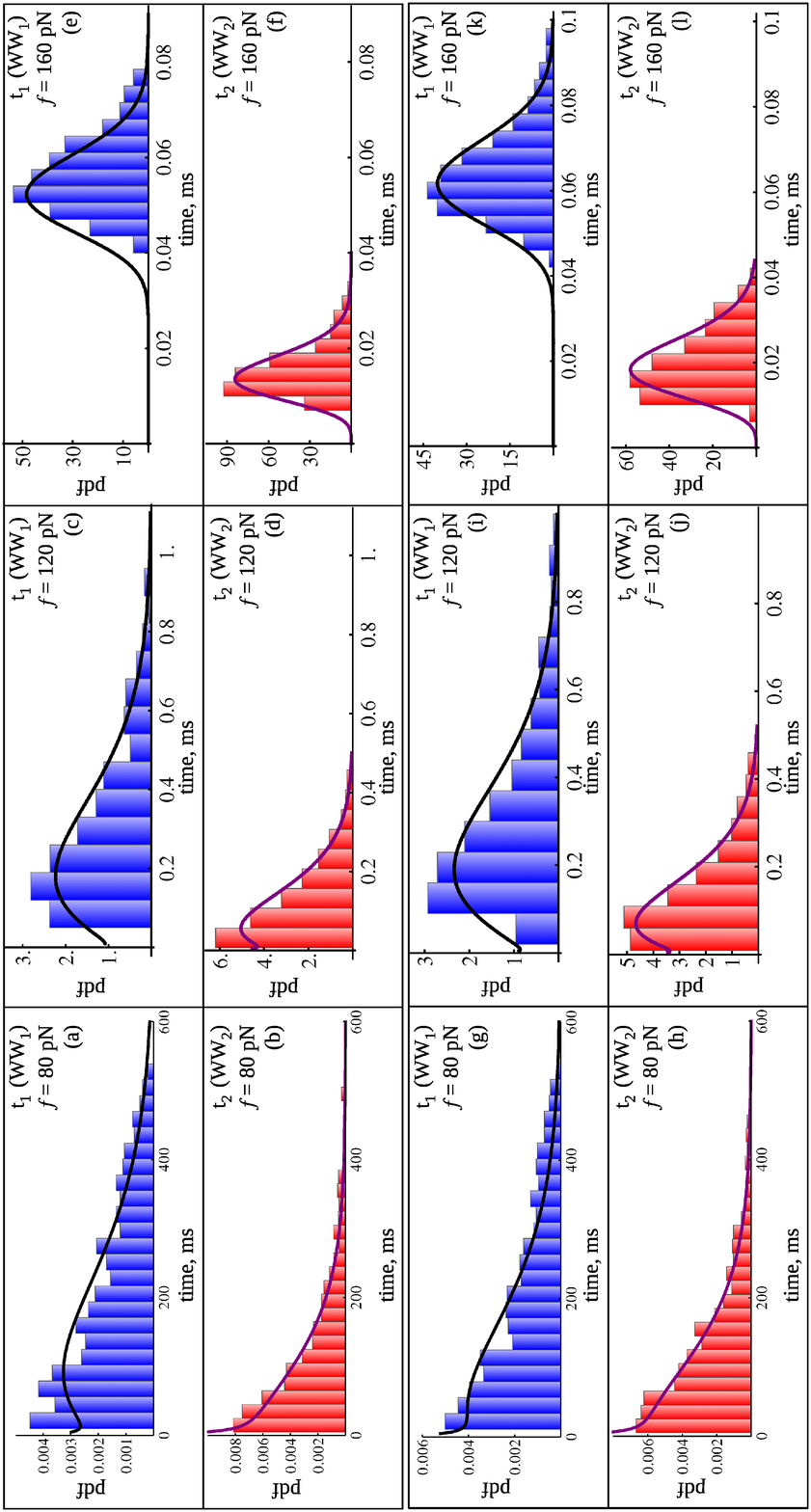}
\caption{{\large(Olga Kononova, Lee Jones, Valeri Barsegov)}}
\end{figure}


\newpage
\begin{figure}[h]\label{fig:figure4}
\includegraphics[width=2.4in]{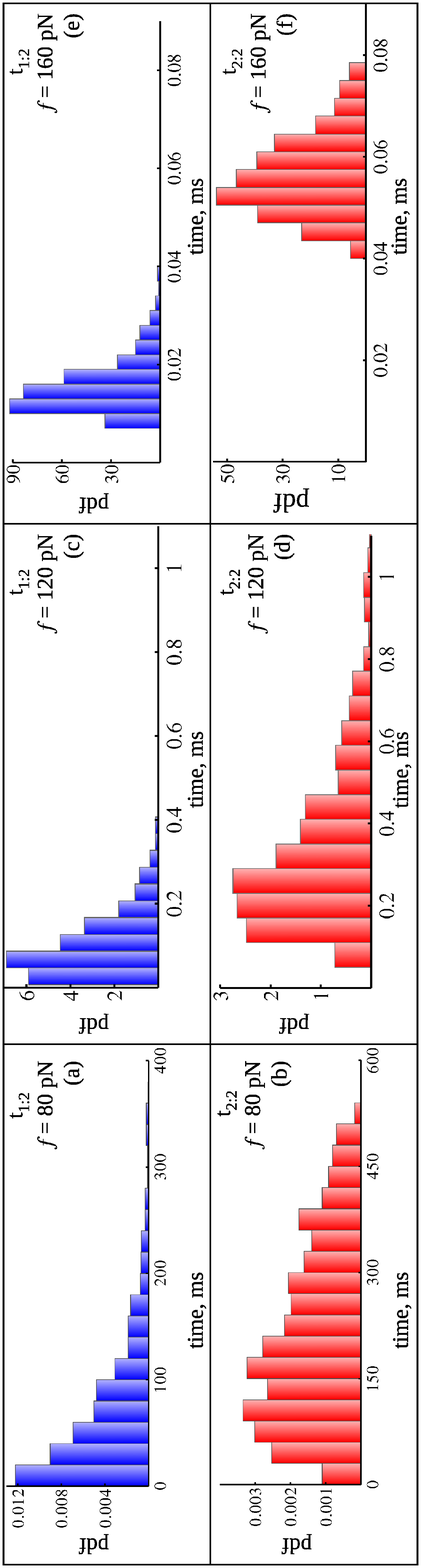}
\caption{{\large(Olga Kononova, Lee Jones, Valeri Barsegov)}}
\end{figure}


\newpage
\begin{figure}[h]\label{fig:figure5}
\includegraphics[width=3.6in]{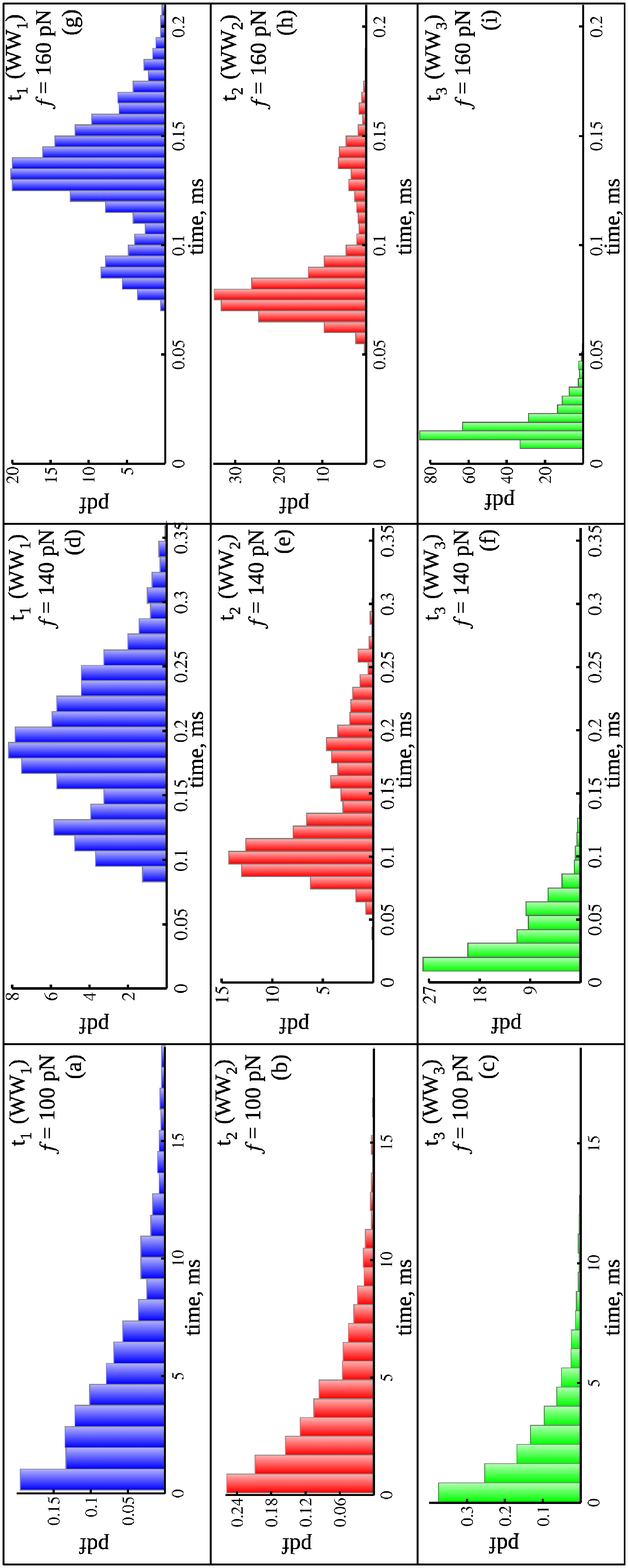}
\caption{{\large(Olga Kononova, Lee Jones, Valeri Barsegov)}}
\end{figure}

    
\end{document}